\documentclass[aps]{article}
\usepackage[top=12mm,bottom=12mm,left=30mm,right=30mm,head=12mm,includeheadfoot]{geometry}
\usepackage{graphicx} 
\usepackage{qcircuit}
\usepackage{physics}
\usepackage{amssymb}
\usepackage{hyperref} 
\usepackage{algorithm}
\usepackage{xcolor}
\usepackage{dsfont}
\usepackage{algpseudocode}
\usepackage{authblk}
\usepackage[parfill]{parskip}
\usepackage[normalem]{ulem}
\usepackage{etoolbox}
\apptocmd{\thebibliography}{\small}{}{}

\newcommand{\cost}{\ensuremath{\mathcal{C}}}

\definecolor{GRAblue}{HTML}{045275} 
\definecolor{GRgreen}{HTML}{089099}
\definecolor{GRlightgreen}{HTML}{7CCBA2}
\definecolor{GRorange}{HTML}{F0746E}
\definecolor{GRpink}{HTML}{DC3977} 

\MakeRobust{\Call}

\title{A simple quantum algorithm to efficiently prepare sparse states}

\author{Debora Ramacciotti\footnote{\href{debora.ramacciotti@itp.uni-hannover.de }{debora.ramacciotti@itp.uni-hannover.de}}}
\author{Andreea I.\  Lefterovici\footnote{\href{andreea.lefterovici@itp.uni-hannover.de}{andreea.lefterovici@itp.uni-hannover.de}}}
\author{Antonio F. Rotundo\footnote{\href{antonio.rotundo@itp.uni-hannover.de}{af.rotundo@gmail.com}}}
\affil{Institut f\"{u}r Theoretische Physik, Leibniz Universit\"{a}t Hannover, Germany}

\date{}

\begin{document}
\maketitle

\begin{abstract}
\noindent State preparation is a fundamental routine in quantum computation, for which many algorithms have been proposed. Among them, perhaps the simplest one is the Grover-Rudolph algorithm. In this paper, we analyse the performance of this algorithm when the state to prepare is sparse. We show that the gate complexity is linear in the number of non-zero amplitudes in the state and quadratic in the number of qubits. We then introduce a simple modification of the algorithm, which makes the dependence on the number of qubits also linear. This is competitive with the best known algorithms for sparse state preparation.
\end{abstract}

\tableofcontents

\section{Introduction}
Given a classical vector $\psi\in \mathbb{C}^N$, the goal of state preparation is to build a unitary $U_\psi$, such that $U_\psi\ket{0}=\ket{\psi}
$\,, where $\ket{\psi}$ is a quantum state whose amplitudes are given by $\psi$. 
This is the first step of many algorithms, such as the
quantum simulation of physical systems \cite{zalka1998simulating, georgescu2014quantum}, quantum machine learning \cite{lloyd2013quantum}, and quantum linear solvers \cite{harrow2009quantum, childs2017quantum}. 
For this reason, state preparation is a subroutine of fundamental importance in quantum computing, and it is an object of ongoing research. 

Early state preparation algorithms are described in \cite{ zalka1998simulating, grover2002creating, kaye2001quantum}. The basic idea of these algorithms is the same: it was independently introduced in \cite{grover2002creating} and \cite{kaye2001quantum}, and had already been present in earlier works such as \cite{zalka1998simulating, knill1995approximation}. Following what is now the standard notation, we collectively refer to these algorithms as Grover-Rudolph. 

In recent years, several works have tried to design new state-preparation algorithms with better worst-case asymptotic scaling, e.g.\ \cite{mottonen2004transformation, sun2023asymptotically, plesch2011quantum}, and have uncovered an interesting trade-off between space and time complexity in state preparation. In this paper, we take a more practical point of view. We focus on \emph{sparse} vectors, i.e.\ vectors with only a few nonzero elements. This is a special class of vectors, which often appears in practical applications, such as quantum linear solvers \cite{harrow2009quantum, childs2017quantum}. 
Recent works that have considered state preparation algorithms tailored for sparse vectors are \cite{gleinig2021efficient, malvetti2021quantum, de2022double, mozafari2022efficient}. 
These algorithms have a complexity linear in both the sparsity of the vector and the number of qubits.

The number of gates required to prepare a generic state with the Grover-Rudolph algorithm scales exponentially in the number of qubits, so this algorithm is sometimes overlooked as an option to prepare sparse states.
Our first contribution is to explicitly show that Grover-Rudolph is able to prepare sparse vectors with a number of gates linear in the sparsity and quadratic in the number of qubits. 
This is a simple result, but, to the knowledge of the authors, it is not clearly stated and proved in the literature.
We then introduce a small modification of the Grover-Rudolph algorithm, which brings down the complexity of preparing sparse vectors to linear in both the sparsity and the number of qubits. 
This shows that Grover-Rudolph is a competitive algorithm for preparing sparse states.

The rest of the paper is organized as follows. In Sec.\ \ref{sec:gr}, we summarize the Grover-Rudolph algorithm. In Sec.\ \ref{sec:sparse-gr}, we specialize to sparse states and analyze the number of gates the Grover-Rudolph algorithm requires for their preparation.
Finally, in Sec.\ \ref{sec:perm_gr}, we introduce a simple variation of Grover-Rudolph algorithm, which we call Permutation Grover-Rudolph, and show that it has the same complexity as more recent algorithms designed for sparse vectors \cite{gleinig2021efficient, malvetti2021quantum, de2022double, mozafari2022efficient}.

\section{Grover-Rudolph algorithm}\label{sec:gr}
In this section, we describe the Grover-Rudolph algorithm \cite{grover2002creating} for state preparation.\footnote{One can consider several versions of the Grover-Rudolph algorithm. The one we present here is similar to the one of \cite{mottonen2004transformation}, except for the use of phase gates, in place of $R_Z$ rotation, and for skipping an optimization step. See the main text for further explanations.} 

Let $\psi\in \mathbb{C}^N$ be a classical vector, we want to implement a unitary $U_\psi$ such that $U_\psi\ket{0}=\ket{\psi}
$\,, where $\ket{\psi}$ is a quantum state with amplitudes equal to $\psi$. 
For simplicity, we assume that $N=2^n$, so that we can encode the vector $\psi$ in a $n$-qubit state.\footnote{If this is not the case, one can pad $\psi$ with zeros until this condition is met.} More precisely, we want that  
\begin{equation}\label{eq:state_to_prepare}
    U_\psi\ket{0}= \frac{e^{i\theta}}{\norm{\psi}}\sum_{i_1\dots i_n}\psi_{i_1\dots i_n}\ket{i_1 \dots i_n},
\end{equation}
where the indices $i_k$ take values in $\{0,1\}$, $\norm{\psi}$ is the 2-norm of the vector, and $\theta\in [0,2\pi)$ is some irrelevant global phase.
The strategy of the Grover-Rudolph algorithm is to construct a series of coarse-grained versions of $\psi$ and prepare them recursively using controlled rotations. 

More precisely, let $\psi^{(k)}$, for $k=1,\dots, n-1$, be the following coarse-grained states with components
\begin{equation}\label{eq:coarse_grained_state}
\psi_{i_1 \dots i_{k}}^{(k)} = 
    \text{arg}(\psi^{(k+1)}_{i_1 \dots i_{k} 0})
    \sqrt{|\psi^{(k+1)}_{i_1 \dots i_{k} 0}|^2 + |\psi^{(k+1)}_{i_1 \dots i_{k} 1}|^2}\,.
\end{equation}
The superscript $(k)$ keeps track of the number of qubits required to encode $\psi^{(k)}$.
For notational convenience, we also introduce $\psi^{(0)}\equiv1$ and $\psi^{(n)}\equiv\psi$.
We prepare states $\ket*{\psi^{(k)}}$, whose amplitudes are given by $\psi^{(k)}$, by recursively appending a qubit in state $\ket{0}$ and performing the following transformation,
\begin{align}\label{eq:recursion}
    \psi_{i_1 \dots i_k}^{(k)}\ket{{i_1 \dots i_k}}\ket{0} \rightarrow \psi_{i_1 \dots i_k}^{(k)}\ket{{i_1 \dots i_k}}  \Bigl(
    \cos{\theta_{i_1 \dots i_k}^{(k)}}\ket{0} +  e^{i\phi^{(k)}_{i_1 \dots i_k}}\sin{\theta_{i_1 \dots i_k}^{(k)}} \ket{1} 
    \Bigr).
\end{align}
The angles and phases should be chosen so that the new state is $\ket*{\psi^{(k+1)}}$, i.e.\ such that the term on the r.h.s.\ of \eqref{eq:recursion} is equal to $\sum_j\psi_{i_1 \dots i_{k}j}^{(k+1)}\ket{i_1 \dots i_{k}j}$.
A short calculation shows that this requires 
\begin{equation}\label{eq:angles_phases}
\begin{split}
    \theta_{i_1 \dots i_k}^{(k)} &= 2 \arccos{\frac{|\psi^{(k+1)}_{i_1 \dots i_{k} 0}|}{|\psi^{(k)}_{i_1 \dots i_k}|}}\,, \\
    \phi_{i_1 \dots i_k}^{(k)} &= \text{arg}(\psi^{(k+1)}_{i_1 \dots i_{k} 1}) - \text{arg}(\psi^{(k+1)}_{i_1 \dots i_{k} 0})\,,
\end{split}
\end{equation}
where $\arg(z)$ is the phase of a complex number $z$, and when $\psi^{(k)}_{i_1 \dots i_k}= 0$ one should pick $\theta_{i_1 \dots i_k}^{(k)}=0$. 
For $k=0$, there are no controlling qubits, so we are simply performing a 1-qubit gate.
The transformation Eq.\ \eqref{eq:recursion} can be implemented by applying a y-rotation $R_y(\theta^{(k)}_{i_1\dots i_k})$
and a phase shift gate $P(\phi^{(k)}_{i_1\dots i_k})$,\footnote{The y-rotation acts on the computational basis as: $R_y(\theta)\ket{0} = \cos(\theta/2) \ket{0} + i\sin(\theta/2)\ket{1}$ and $R_y(\theta)\ket{1} = \cos(\theta/2) \ket{0} - i\sin(\theta/2)\ket{1}$. The phase shift gate acts on the computational basis as: $P(\phi)\ket{0} = \ket{0}$ and $P(\phi)\ket{1} = e^{i\phi}\ket{1}$.} 
both controlled on the state of the first $k$ qubits being $\ket{i_1\dots i_k}$,
\begin{equation}\label{eq:layer_k}
    U_k = \sum_{i_1\dots i_{k}}\ketbra{i_1\dots i_{k}}\otimes\bigl(P(\phi^{(k)}_{i_1\dots i_k})\cdot R_y(\theta^{(k)}_{i_1\dots i_k}) \bigr)\,,\quad k=0,\dots, n-1\,.
\end{equation}
Notice that the superscripts of the angles and phases indicate how many qubits control the transformation, and the subscripts which value the controls should have.
For instance, $\theta^{(2)}_{11}$ means that the rotation and phase gates are applied when the first two qubits are in state $\ket{11}$.

The steps we have just explained are summarized in Alg.\ \ref{alg:gr}. The algorithm takes as input the angles and phases needed to implement the $U_k$'s. 
We decide to store these in a list of dictionaries, $L_k$ for $k=0, \dots, n-1$.
The entries in the dictionary $L_k$ are given by $\{\text{key: value}\}$ pairs of the form $\{(i_1,\dots, i_k)\text{: }(\theta^{(k)}_{i_1\dots i_k}, \phi^{(k)}_{i_1\dots i_k})\}$. 
For the special case $k=0$, we set $L_0=\{1\text{: }(\theta^{(0)}, \phi^{(0)})\}$.
This dictionary can be computed using Alg.\ \ref{alg:findAngles}.

\begin{algorithm}
\caption{Grover-Rudolph}\label{alg:gr}
\begin{algorithmic}[1]
\Function{GroverRudolph}{angles and phases dictionaries $L_k$}
    \State $\ket{\psi} \gets 1$
    \For{$k=0, \dots, n-1$}
    \State $\ket{a}\gets \ket{0}$
    \For{$(i_1, \dots, i_k), (\theta^{(k)}_{i_1\dots i_k}$, $\phi^{(k)}_{i_1\dots i_k})$ \textbf{in} $L_k$}
    \Comment{Implement $U_k$ as in Eq.\ \eqref{eq:layer_k}}
    \If{$\ket{\psi}=\ket{i_1, \dots, i_k}$ }
    \State $\ket{a} \gets  P(\phi^{(k)}_{i_1\dots i_k}) \cdot R_y(\theta^{(k)}_{i_1\dots i_k})\ket{a}$
    \EndIf
    \EndFor
    \State $\ket{\psi}\gets \ket{\psi}\otimes \ket{a}$
    \EndFor
    \State \textbf{return} $\ket{\psi}$
\EndFunction
\end{algorithmic}
\end{algorithm}

\begin{algorithm}
\caption{FindAngles}\label{alg:findAngles}
\begin{algorithmic}[1]
\Function{FindAngles}{$\psi\in\mathbb{C}^N$ with $N=2^n$}
\State $\psi^{(n)}\gets \psi$
\For{$k=n-1, n-2,\dots, 1$}
    \State $L_k$ $\gets$ empty dictionary
    \State Compute $\psi^{(k)}$ using Eq.\ \eqref{eq:coarse_grained_state}
    \For{$(i_1, \dots, i_k)\in \{0,1\}^{k}$}
    \State Compute $\theta^{(k)}_{i_1\dots i_k}$ and $\phi^{(k)}_{i_1\dots i_k}$ using Eq.\ \eqref{eq:angles_phases}
    \State $L_k[(i_1, \dots, i_k)] \gets (\theta^{(k)}_{i_1\dots i_k}, \phi^{(k)}_{i_1\dots i_k})$ 
    \EndFor
\EndFor
\State \textbf{return} $\{L_k\}$
\EndFunction
\end{algorithmic}
\end{algorithm}

We now analyze the complexity of both Alg.\ \ref{alg:gr} and Alg.\ \ref{alg:findAngles}. 
Let's begin with the quantum part, Alg.\ \ref{alg:gr}. 
The algorithm consists of $n$ unitaries, $U_k$ for $k=0,1,\dots, n-1$. 
Each unitary involves $k+1$ qubits and is made out of  $R_y$ rotations and phase gates controlled on $k$ qubits ($U_0$ is not controlled). 
In the worst-case, the $k$-th unitary is made out of $O(2^k)$ controlled gates. For each controlled gate, we need $O(k)$ Toffoli gates. Summing over all $k$, we arrive at a worst-case asymptotic scaling of order $O(n2^n)$.
The complexity of the classical preprocessing required to determine the rotation angles and the phases is $O(2^n)$. 
To see this, consider the function FindAngles in Alg.\ \ref{alg:findAngles}. 
To find the angles, we need first to find the coarse-grained states $\psi^{(k)}$.
In the worst case, $\psi^{(k)}$ has $2^k$ nonzero components, so to find the next coarse-grained state we need $O(2^k)$ operations.
In total, we need $O(2^n)$ operations to find all coarse-grained states.
To find the angles, we need the same amount of operations. 

Notice that one could use the efficient gate decomposition from \cite{mottonen2004transformation} to bring the worst-case gate complexity down to $O(2^n)$.
We don't do this because when specializing to sparse vectors, most angles and phases are zero.
The number of multi-controlled 1-qubit gates is then much smaller than in the worst case scenario, and the decomposition of \cite{mottonen2004transformation} would in fact lead to much deeper circuits.\footnote{To be more explicit, consider Figures 1 and 2 from \cite{mottonen2004transformation}. The decomposition proposed there works by replacing the multi-controlled rotations, with $2^k$ CNOTs and $2^k$ 1-qubit rotations, with angles defined by Eq.\ (3) in \cite{mottonen2004transformation}. For sparse vectors, it turns out that most angles on the r.h.s.\ of Eq.\ (3) are zero. The angles on the l.h.s.\ on the other hand are typically all different from zero. So for sparse vectors, we actually end up with a less efficient circuit.}

\subsection*{A simple example}\label{sec:example}
\begin{figure}[h!]
    \centering
    \includegraphics[scale=0.7]{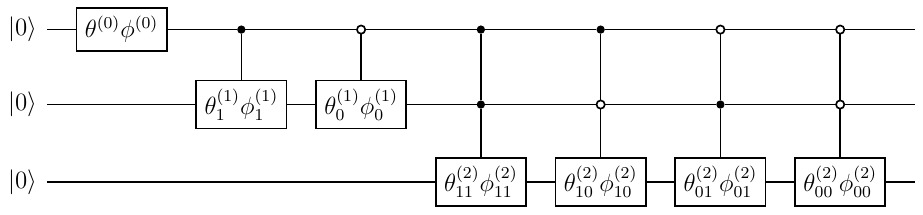}
    \caption{The general Grover-Rudolph circuit for preparing a 3-qubit state.}
    \label{fig:grcircuit}
\end{figure}

Before continuing, we illustrate the Grover-Rudolph algorithm in a simple example. 
Consider a vector with $8$ positive components
\begin{equation}
    \psi= \begin{bmatrix}
        0& \sqrt{\frac{1}{3}}& 0& 0& 0& 0& \sqrt{\frac{2}{3}}& 0
        \end{bmatrix},
\end{equation}
our goal is to prepare the corresponding quantum state
\begin{equation}
    \ket{\psi} = \sqrt{\frac{1}{3}}\ket{001} + \sqrt{\frac{2}{3}}\ket{110}.
\end{equation}
The most general circuit implementing Alg.\ \ref{alg:gr} for 3 qubits is depicted in Fig.\ \ref{fig:grcircuit}.
Note that to shorten the notation, we have denoted the rotation gates with $\theta$ instead of $R_y(\theta)$, and phase gates with $\phi$ instead of $P(\phi)$. 

\begin{figure}
    \centering
    \includegraphics[scale=0.75]{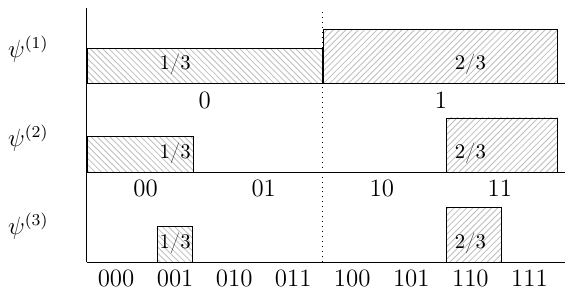}
    \caption{Visualization of coarse-graining procedure from Eq.\ \eqref{eq:coarse_grained_state}. The bars in the histograms correspond to probabilities, i.e.\ amplitudes squared. Note that $\psi^{(3)} \equiv \psi$.}
    \label{fig:Grudolph_probabilities}
\end{figure}

To find the angles and phases, we first need to compute the coarse-grained vectors, as in Eq.\ \eqref{eq:coarse_grained_state}. 
Since the vector $\psi$ is positive, we can interpret its entries as the square root of a probability, and 
visualize the coarse-graining procedure as in Fig.\ \ref{fig:Grudolph_probabilities}. 
The coarse-grained vectors $\psi^{(k)}$ are obtained by iteratively binning together the probabilities in pairs and summing them. 

The Grover-Rudolph procedure starts by preparing the first coarse-grained state 
\begin{equation}
    \ket*{\psi^{(1)}} = \sqrt{\frac{1}{3}}\ket{0}+\sqrt{\frac{2}{3}}\ket{1}. 
\end{equation}
by applying a 1-qubit rotation to the 
$\ket{0}$. 
This can be done by applying $R_y(\theta^{(0)})$ with $\theta^{(0)}=2\cos\!{(1/\sqrt{3})}$.
We can then prepare the next coarse-grained state, 
\begin{equation}
    \ket*{\psi^{(2)}} = \sqrt{\frac{1}{3}}\ket{00}+\sqrt{\frac{2}{3}}\ket{11}, 
\end{equation}
by appending a qubit in state $\ket{0}$, and rotating it depending on the state of the first qubit. 
Namely, when the first qubit is in state $\ket{1}$, we need to rotate the second to $\ket{1}$; when the first qubit is in state $\ket{0}$, we should leave the second qubit in $\ket{0}$.
This can be done by picking $\theta^{(1)}_{0}=0$ and $\theta^{(1)}_{1}=\pi$.
The last step is performed similarly.
We need to pick angles $\theta^{(2)}_{00}=\pi$ and $\theta^{(2)}_{11}=0$.
The end of this process yields the desired state $\ket{\psi}$.

\section{Grover-Rudolph for sparse vectors}\label{sec:sparse-gr}
In this section, we analyze how well Grover-Rudolph performs for preparing sparse vectors.
We consider vectors $\psi\in\mathbb{C}^N$ which have only $d\ll N$ nonzero elements.
Above, we have seen that the worst-case complexity of Grover-Rudolph is exponential in the number of qubits.
However, we intuitively expect that for sparse vectors it should be possible to prepare the state with only $O(d)$ gates, as the vector has only $d$ degrees of freedom.
Below, we explicitly show this and find that Grover-Rudolph can prepare sparse states with $O(dn^2)$ gates. 

We assume that we know the number of nonzero elements in $\psi$ and their locations. Namely, that we have access to $\psi$ as a tuple of vectors, $(\lambda, \phi)$.
The vector $\lambda$ contains the locations of the nonzero entries of $\psi$, and $\phi$ contains their values.
The length of both vectors is $d$.
Without loss of generality, we assume that the elements of $\lambda$ are arranged in increasing order. 

The coarse-graining procedure from Eq.\ \eqref{eq:coarse_grained_state} doesn't increase the number of nonzero elements, hence all $\psi^{(k)}$ have sparsity $d_k\le d$. 
Let $(\lambda^{(k)}, \phi^{(k)})$ be the pair of vectors defining $\psi^{(k)}$.
Eq.\ \eqref{eq:coarse_grained_state} needs to be applied only when two consecutive elements of $\psi$ are nonzero.
From this follows that we can find each coarse-grained vector using at worst $O(d)$ operations, for a total of $O(dn)$ classical operations.
Once we have these vectors, we can find the angles and phases with additional $O(dn)$ operations. 
In Alg.\ \ref{alg:sparseAngles}, we provide the pseudocode for finding angles exploiting the sparsity of the vector.
Notice that this is the only thing we need to change to take advantage of sparsity, Alg.\ \ref{alg:gr} remains unchanged.

Since each $\psi^{(k)}$ has at most $d$ nonzero elements, we find that each $U_k$ has at most $d$ nonzero angles and phases. From this it follows that each $U_k$ can be implemented using $O(kd)$ gates ($d$ from the number of nonzero angles, $k$ from the number of controlling qubits). 
Summing over $k$, we arrive at an overall gate complexity of $O(dn^2)$.
As expected, we conclude that Alg.\ \ref{alg:gr} performs on sparse vectors much better than the worst-case complexity would suggest. 

\begin{algorithm}[H]
\caption{FindSparseAngles}\label{alg:sparseAngles}
\begin{algorithmic}[1]
\Function{FindSparseAngles}{nonzero location and values $(\lambda, \phi)$}
\State $(\lambda^{(n)}, \phi^{(n)})\gets(\lambda, \phi)$
\For{$k=n-1, n-2,\dots, 1$}
    \State $\lambda^{(k)}\gets [\;]$, $\phi^{(k)}\gets [\;]$ 
    \State $L_k\gets \{\;\}$
    \Comment{Empty dictionary for angles and phases of unitary $U_k$}
    \For{$l=0,1,\dots, \text{len}(\lambda^{(k+1)})$}
    \If{$\lambda_l^{(k+1)}$ is even and $\lambda_{l+1}^{(k+1)}=\lambda_{l}^{(k+1)}+1$}
    \State $x \gets$ evaluate Eq.\ \eqref{eq:coarse_grained_state}
    \Comment{Use $\psi^{(k+1)}_{i_1 \dots i_{k} j}\gets\phi^{(k+1)}_{l+j}$}
    \State Append $x$ to $\phi^{(k)}$
    \State Append $\lambda_l^{(k+1)}/2$ to $\lambda^{(k)}$
    \State $l\gets l+1$
    \Comment{Skip one iteration}
    \Else
    \State Append $\phi^{(k+1)}_{l}$ to $\phi^{(k)}$
    \State Append $\lfloor\lambda_l^{(k+1)}/2\rfloor$ to $\lambda^{(k)}$
    \EndIf
    \State Compute $\theta^{(k)}_{i_1\dots i_k}$ and $\phi^{(k)}_{i_1\dots i_k}$ using Eq.\ \eqref{eq:angles_phases}
    \Comment{Use $\psi^{(k)}_{i_1 \dots i_{k}}\gets x$}
    \State $L_k[(i_1, \dots, i_k)] \gets (\theta^{(k)}_{i_1\dots i_k}, \phi^{(k)}_{i_1\dots i_k})$ 
    \Comment{$(i_1, \dots, i_k)$ bit representation of $\lfloor\lambda_l^{(k+1)}/2\rfloor$}
    \EndFor
\EndFor
\State \textbf{return} $\{L_k\}$
\EndFunction
\end{algorithmic}
\end{algorithm}

\begin{figure}
    \centering
\includegraphics[width=0.48\textwidth]{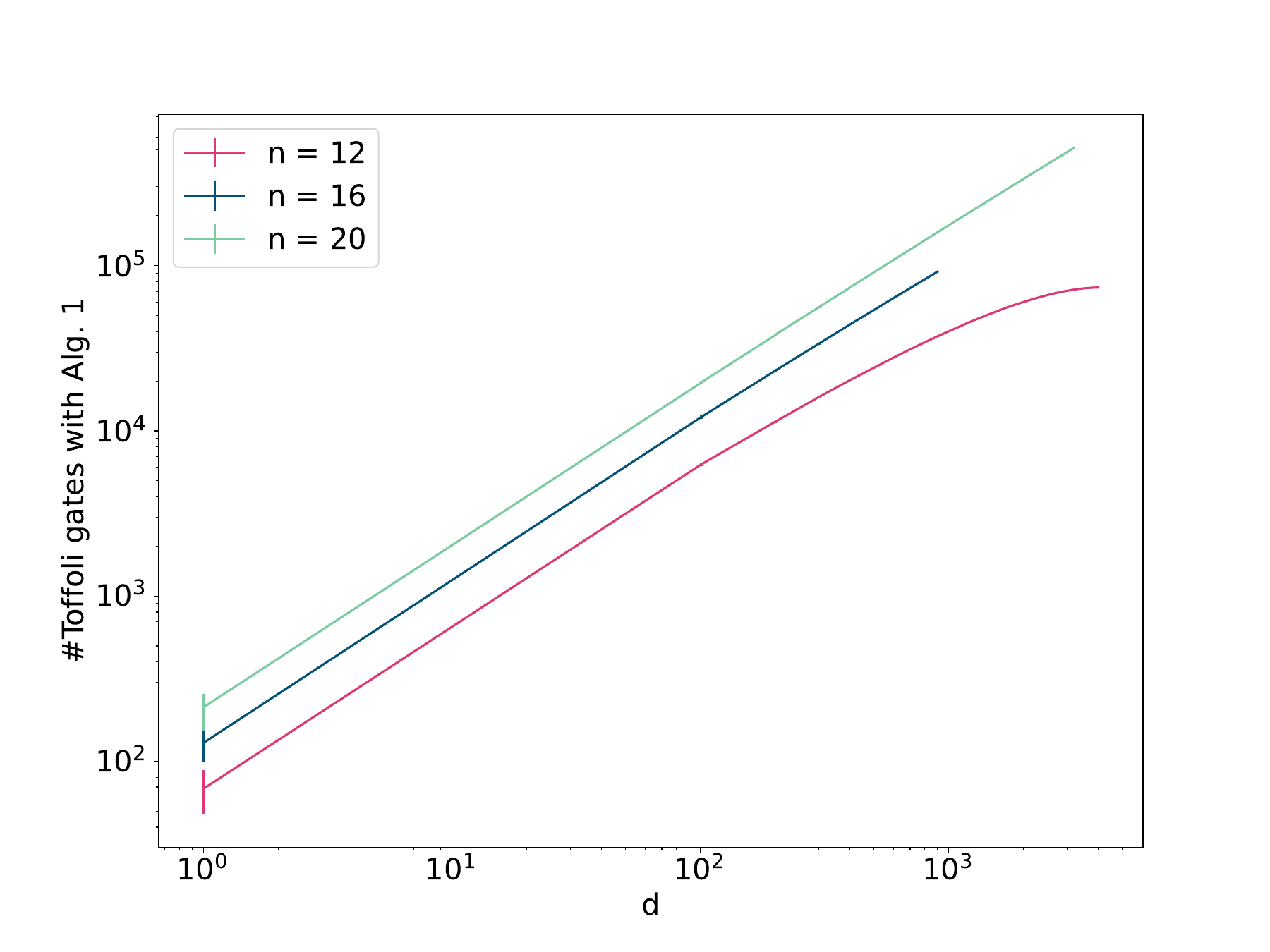}
\includegraphics[width=0.48\textwidth]{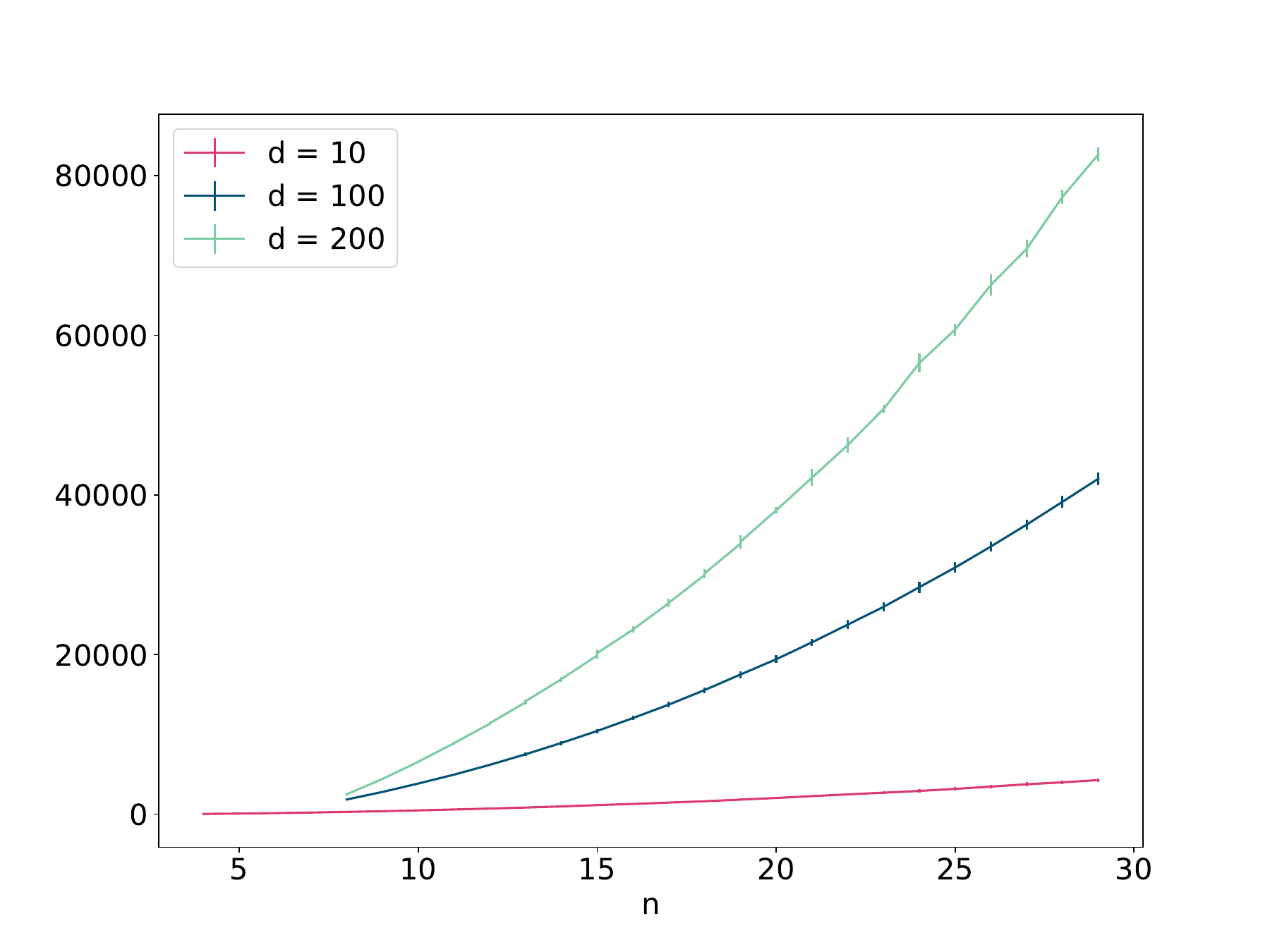}
\caption{Gate count for preparing random states using Alg.\ \ref{alg:gr}, as a function of $d$ at fixed $n$ (a), and as a function of $n$ at fixed $d$ (b). We use a log-log scale in (a) and a lin-lin in (b).} 
    \label{fig:sparseGR}
\end{figure}

On practical instances, the performance of Grover-Rudolph might very well be better than the worst-case.
So, we estimate the typical complexity of the algorithm by sampling random sparse vectors and explicitly counting the gates needed to prepare them.
We compute the cost of the algorithm by counting the number of Toffoli, CNOT, and 1-qubit gates needed to implement it.
We use standard constructions for controlled gates, see e.g.\ \cite{nielsen2010quantum}.
To implement a 1-qubit gate controlled on $k\ge 2$ qubits being in a state given by the bit string $x$, we use $k-1$ ancilla qubits, $2(k-1)$ Toffoli gates, 2 CNOT's, and $4+2(n-\abs{x})$ 1-qubit gates.
Here $\abs{x}$ is the Hamming weight of the bit string $x$.

In more detail: we consider 100 random complex vectors, for various values of $d$ and $n$, and we study how the gate count scales as a function of $d$ and $n$. 
The results are displayed in Fig.\ \ref{fig:sparseGR}.
Notice that we display only the count of Toffoli gates, as the plots for the count of CNOT and 1-qubit gates are very similar.
By inspecting the diagram, we find that also the average-case complexity of the algorithm scales linearly in $d$ and quadratically in $n$.

In this section, we have shown that the Grover-Rudolph algorithm works well on sparse vectors.
However, the scaling we have found, $O(dn^2)$, doesn't quite match the best known algorithms for preparing sparse vectors \cite{gleinig2021efficient, malvetti2021quantum, de2022double, mozafari2022efficient}, which have a worst-case complexity of order $O(dn)$.
Notice that $n$ is only logarithmic in the size of the vector, so it is possible that optimizing the Grover-Rudolph circuit might overcome this small overhead in practical applications. 
In App.\ \ref{app:optimization}, we consider a simple optimization procedure that reduces the number of needed gates, at the cost of a small classical overhead. 
However, we find that this reduction is not significant enough.
Therefore, in the next section we introduce a small variant of Grover-Rudolph for which we can prove a worst-case scaling of $O(dn)$.

\section{Permutation Grover-Rudolph}\label{sec:perm_gr}
We introduce a simple variation of Grover-Rudolph, for which we can prove the worst case complexity of order $O(dn)$. The idea of this variant is as follows. First, we prepare a dense state whose amplitudes are given by the nonzero entries of $\psi$. To prepare this state, we only need $\lceil \log{d}\rceil$ qubits, and we can use Alg.\ \ref{alg:gr}. We then append a sufficient number of qubits, such that the total dimension of the Hilbert space becomes $N$, and apply a permutation unitary which maps the nonzero amplitudes to their correct location. We show that this permutation can be efficiently implemented. 
The idea of preparing a dense state with all the nonzero entries and then permuting the basis states has already been used in \cite{malvetti2021quantum}.
Our algorithm differs in the implementation of the permutation and, as we discuss further below, has a better classical complexity.

Similarly to the previous section, we assume we have access to $\psi$ as a tuple of vectors $(\lambda, \phi)$. Each vector has size $d$, with $\lambda_i \in \{0, \dots, N\}$ being the location of the $i$-th nonzero element of $\psi$, and $\phi_i$ being its value. 
We assume without loss of generality that the elements of $\lambda$ are arranged in increasing order. As a first step, we prepare a dense vector
\begin{equation}
    \ket*{\tilde{\psi}}=\sum_{i=0}^{d-1} \phi_i\ket{i}\,,
\end{equation}
using standard Grover-Rudolph. 
We then add a sufficient number of qubits initialized in $\ket{0}$, such that the total size of the Hilbert space becomes $N$. 
Finally, we apply a permutation unitary that maps $\ket{i}\rightarrow \ket{\lambda_i}$.

There are of course many permutations mapping $i$ to $\lambda_i$. We build one, directly decomposed in cycles, as follows.
Let $i\in I$, with $I=\{0,1,\dots, d-1\}$, and for simplicity, consider $i=0$. We initialize a cycle with elements $(0, \lambda_0)$. If $\lambda_0\ge d$, we can close the cycle, add it to our permutation, and go to the next available $i$. If $\lambda_0< d$, we set $j= \lambda_0$, we remove $j$ from $I$, and we add $\lambda_j$ to the cycle. We then repeat the steps above until we find a $\lambda_j$ larger than $d$. The steps are summarized in Alg. \ref{alg:sparsePerm}.
\begin{algorithm}
\caption{Sparse Permutation}\label{alg:sparsePerm}
\begin{algorithmic}
\Function{SparsePerm}{vector of nonzero locations $\lambda$}
\State $P \gets \{\,\}$
\Comment{Initialize empty permutation}
\State $S=\{S_0, S_1,\dots, S_{d-1}\}$ with $S_i=1$, for $i=0,\dots, d-1$
\Comment{Track valid starting values for cycles}
\For{$i=0,1,\dots, d-1$}
\If{$S_i=0$ or $\lambda_i = i$}
\State \textbf{continue}
\Comment{Skip iteration if $i$ already present in a cycle or if cycle is trivial}
\EndIf
\State $j \gets \lambda_i$\,, $c\gets \{i, j\}$
\While{$j < d$}
\State $S_j\gets 0$, $j \gets \lambda_{j}$
\State append $j$ to $c$
\EndWhile
\State append $c$ to $C$
\Comment{Add cycle to permutation}
\EndFor
\State \Return $P$
\Comment{Permutation decomposed in cycles}
\EndFunction
\end{algorithmic}
\end{algorithm}

To understand how the complexity of Alg.\ \ref{alg:sparsePerm} scales, let $P=\{c_0, c_1,\dots c_{n_c-1}\}$ be the list of cycles returned by the algorithm. We denote by $M_k$ the length of the $k$-th cycle in the list, with $k=0,1,\dots, n_c-1$. 
To generate this list, the algorithms loops over $i=0,1,\dots, d-1$ and does 3 blocks of operations: an if statement, an initialization, and a while loop. The first two operations take constant time, both $O(n)$. 
The if statement is run every iteration, and the initialization step is run $n_c\le d$ times.
So they both contribute $O(dn)$ to the classical complexity. 
The while loop is run only for cycles with more than 2 entries, with each iteration taking $O(n)$ time. Hence, we find a contribution of order $O\bigl(n\sum_k \max(M_k-2,0)\bigr)$. 
We can upper bound the sum by $\sum_k M_k$, which is the total length of the cycles in $P$. 
In the worst case, we have $\sum_i M_i=2d$, which happens for a permutation made of $d$ 2-cycles. This happens when $\lambda_i\ge d$ for all $i$. 
To see this, let $\lambda_j<d$ for some value of $j$. Then in the permutation we replace two cycles of length 2 with one cycle of length 3, and $\sum_i M_i$ decreases. 
We conclude that Alg.\ \ref{alg:sparsePerm} has complexity of order $O(nd)$.

Once we have decomposed the permutation in cycles, we can use Alg.\ \ref{alg:cycle} (see App.\ \ref{sec:permutations}) to implement it. Alg.\ \ref{alg:cycle} implements a cycle of length $M$ with $O(Mn)$ classical operations. 
Notice that since, as we have argued above, $\sum_i{M_i}=O(d)$, the overall classical complexity is still of order $O(nd)$.
Putting everything together, we find Alg.\ \ref{alg:gr-perm} for preparing sparse states. 
\begin{algorithm}[H]
\caption{Permutation Grover-Rudolph}\label{alg:gr-perm}
\begin{algorithmic}
\Function{PermGR}{sparse vector $\{(x_0, \psi_0), \dots, (x_{d-1}, \psi_{d-1})\}$, number of qubits $n$}
\State Apply Alg.\ \ref{alg:gr} to prepare $\ket*{\tilde{\psi}}=\sum_{i=0}^{d-1} \psi_i\ket{i}$
\State Append $n-\bigl\lceil \log_2{d}\bigr\rceil$  qubits in state $\ket{0}$
\Comment{Add qubits until there are $n$}
\State $P \gets$ \Call{SparsePerm}{$\{x_i\}$}
\For{$c\in P$}
\State Apply \Call{Cycle}{$c$, $n$}
\Comment{See Alg.\ \ref{alg:cycle}}
\EndFor
\EndFunction
\end{algorithmic}
\end{algorithm}

The quantum complexity of this algorithm is given by the cost of the Grover-Rudolph step needed to prepare $\tilde{\psi}$, and the cost of implementing the permutation $\ket{i}\rightarrow \ket{\lambda_i}$. As explained in Sec.\ \ref{sec:gr}, the first is given by $O(n 2^n)$, where $n$ is the number of qubits. For the Grover-Rudolph step in this case, the number of qubits is only $n=O(\log d)$, hence we find $O(d\log d)$. For the second one, we can use Eq.\ \eqref{eq:cost_perm}, which states that the complexity of one cycle scales like $O(Mn)$, where $n$ is the number of qubits and M is the cycle length. 
The cost of the permutation scales like the sum of all cycles lengths times the number of qubits. 
In the worst case, we have $d$ cycles of length 2, obtaining that the complexity of the permutation is $O(dn)$. 
Putting everything together, we find that the worst-case complexity of the algorithm scales as $O(dn)$.
Notice that we could consider better algorithms for the Grover-Rudolph step, e.g.\ the algorithm of \cite{mottonen2004transformation}, which would scale as $O(d)$ instead of $O(d\log d)$. Since the complexity of the algorithm is dominated by the permutation step, we don't think this would make a significant difference.
It would be interesting to consider better ways to implement the permutation.

\begin{figure}
    \centering
\includegraphics[width=0.48\textwidth]{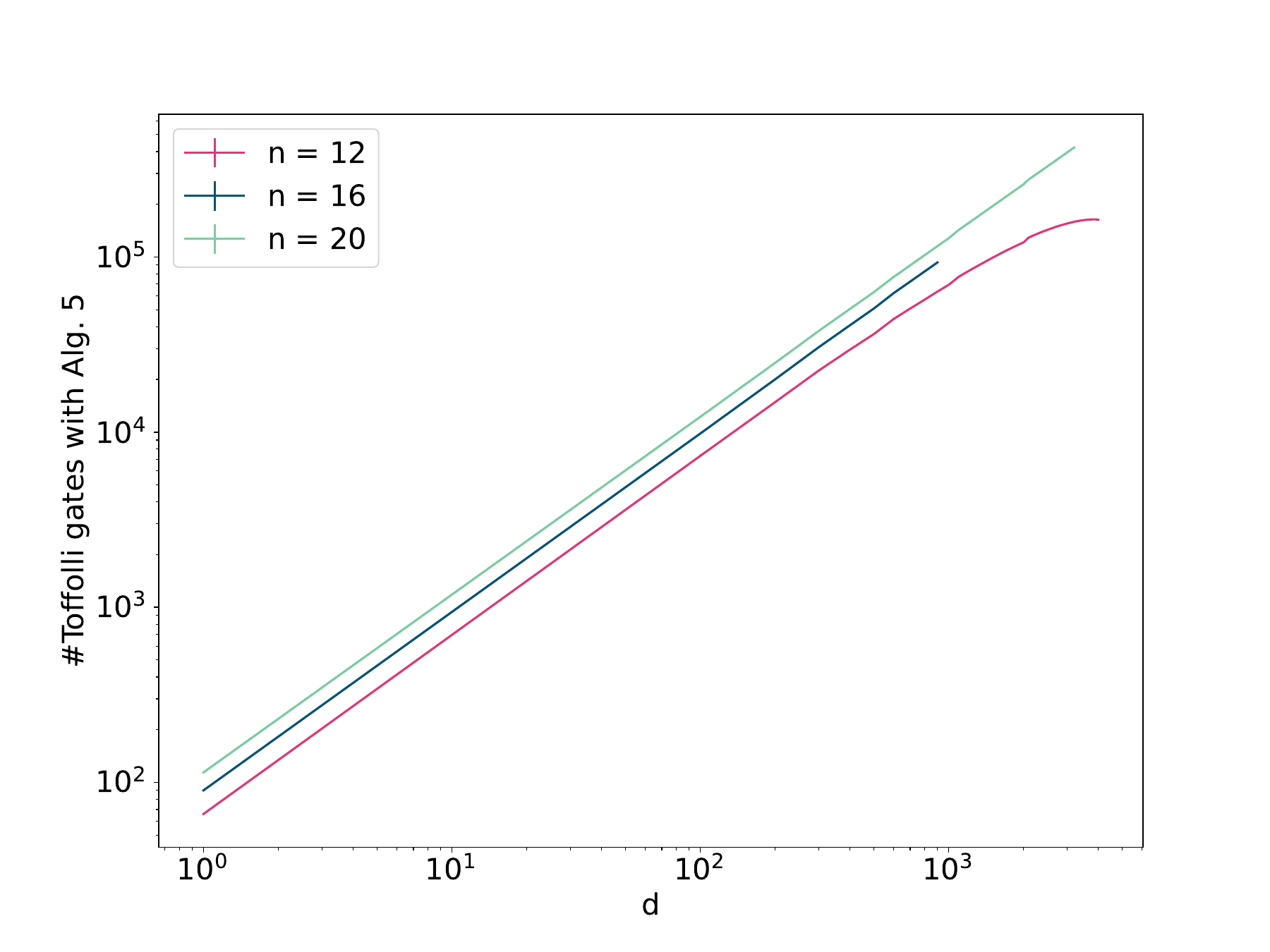}
\includegraphics[width=0.48\textwidth]{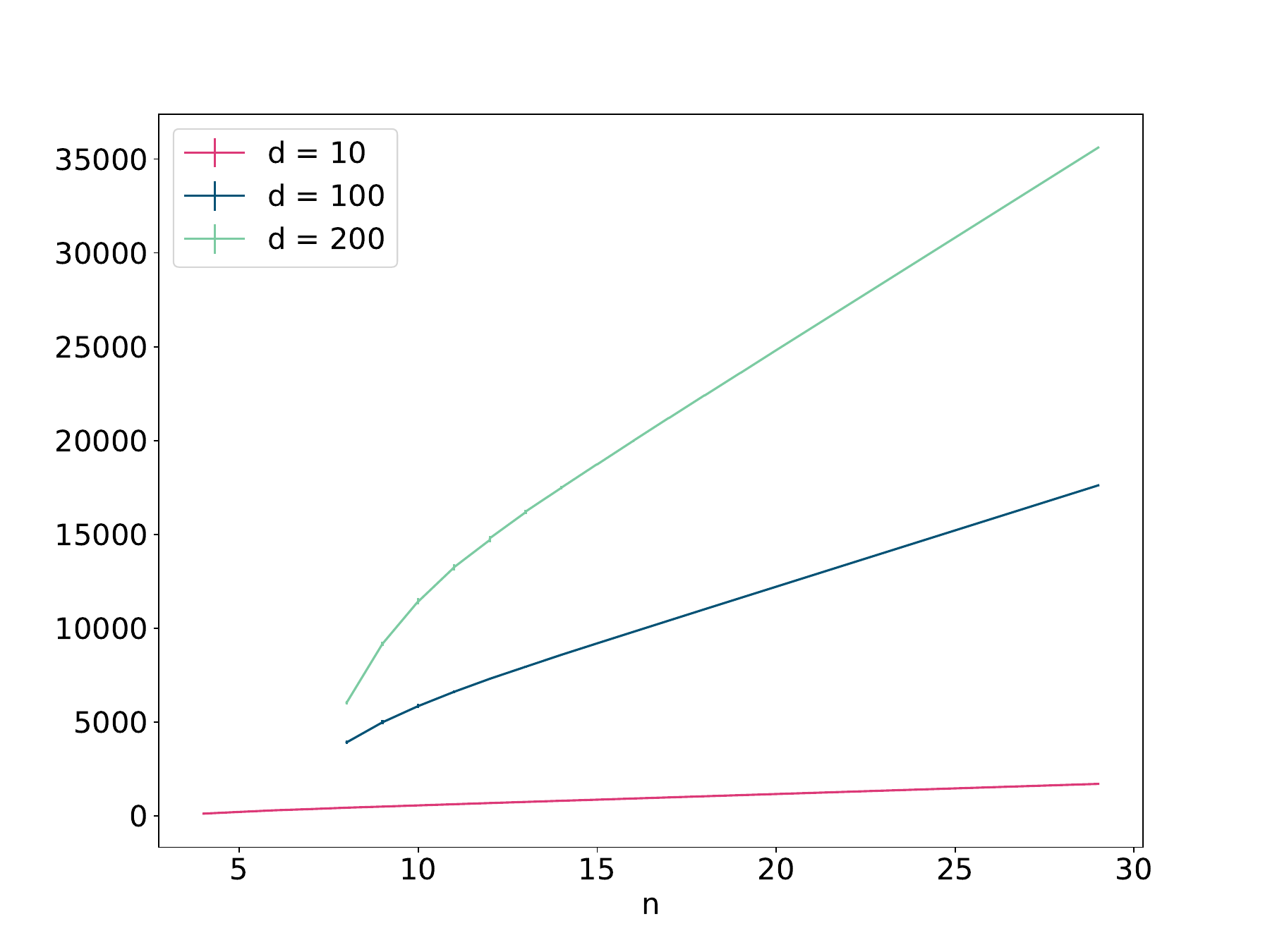}
    \caption{Gate count for preparing random states using Alg.\ \ref{alg:gr-perm}, as a function of $d$ at fixed $n$ (a), and as a function of $N = 2^n$ at fixed $d$ (b). We use a log-log scale in (a) and a lin-lin in (b).} 
    \label{fig:permGR}
\end{figure}

Similarly to what we did in Sec.\ \ref{sec:sparse-gr}, we numerically estimate the average-case complexity of Alg.\ \ref{alg:gr-perm}.
We consider random complex sparse vectors, for various values of $d$ and $n$, and compute the number of gates required by Alg.\ \ref{alg:gr-perm} to prepare them.
The results are shown in Fig.\ \ref{fig:permGR}.
We find that scaling is linear in both $d$ and $n$, the same as in the worst-case analysis.

\begin{figure}[h!]
    \centering
\includegraphics[width=0.48\textwidth]{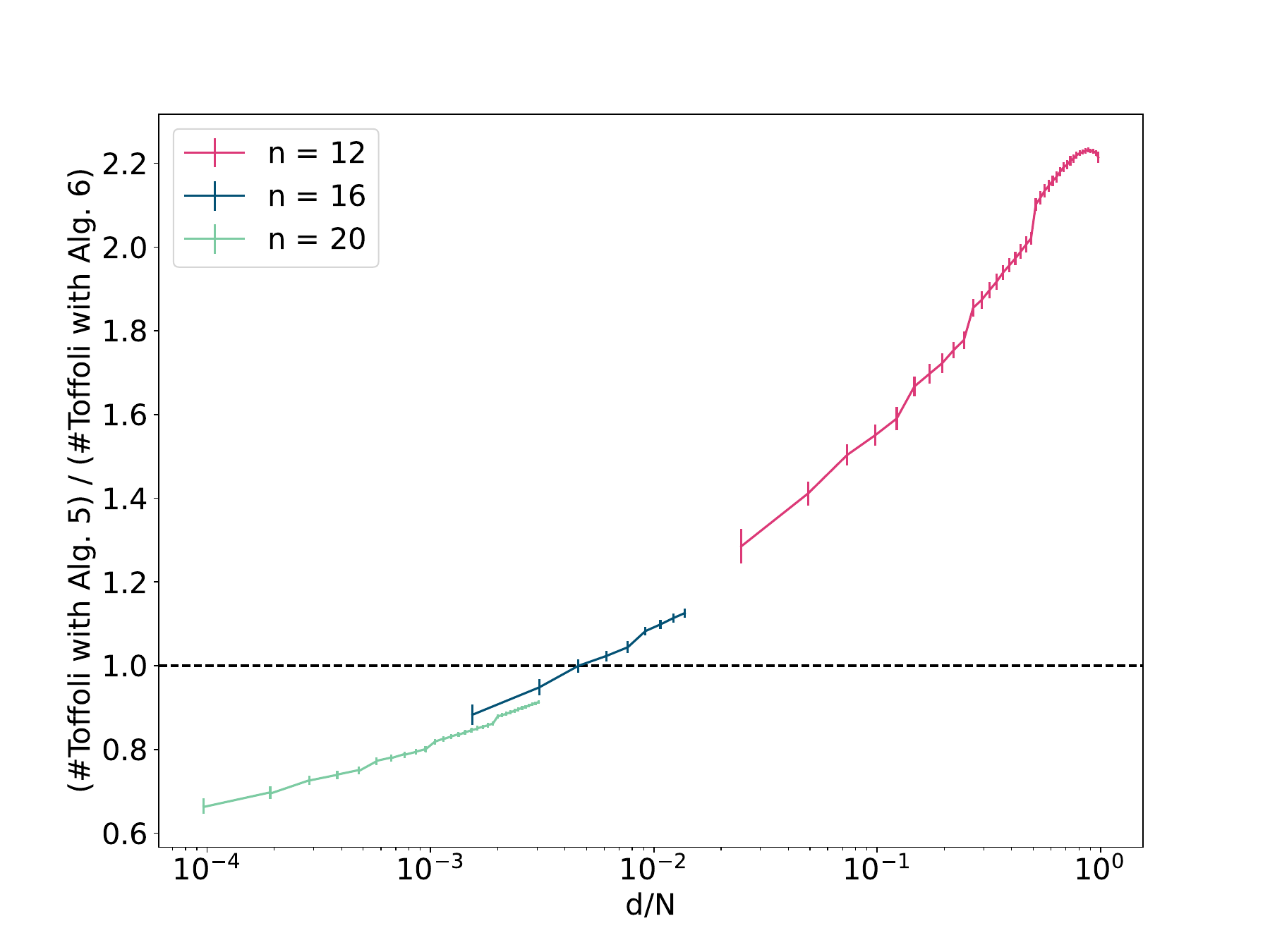} 
\includegraphics[width=0.48\textwidth]{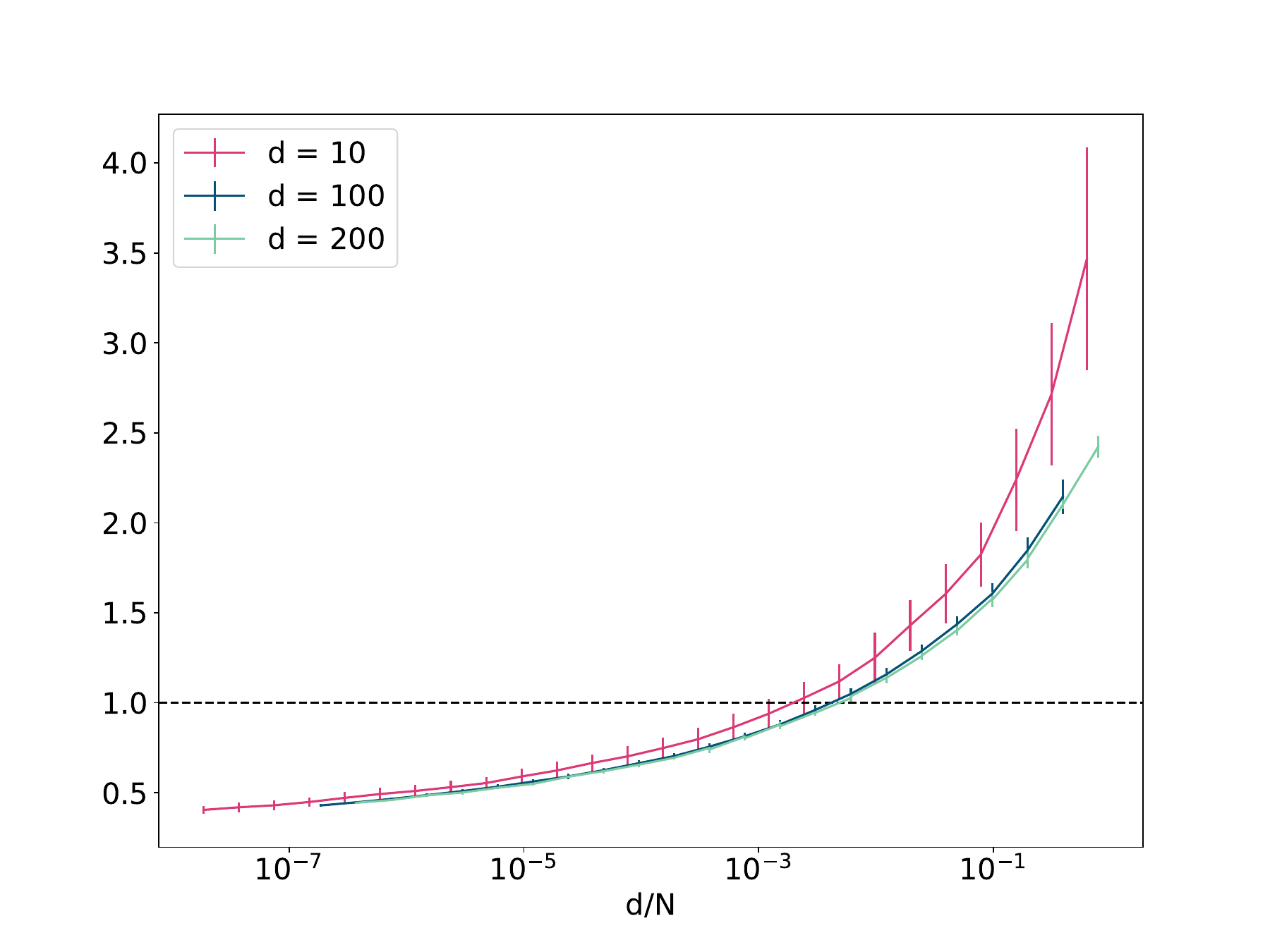}
    \caption{Ratio between the number of gates required by Alg.\ \ref{alg:gr-perm} and the optimized gate count given by Alg.\ \ref{alg:gr} in concert with Alg.\ \ref{alg:optimize_angles} to prepare some random states,  as a function of the density $d/N$ at fixed $n$ (a), and at fixed $d$ (b). We use a log scale on the abscissa.}
\label{fig:eval_permGR}
\end{figure}

From our analysis, we know that the worst-case cost of Alg.\ \ref{alg:gr-perm} scales better than that of Alg.\ \ref{alg:gr}. 
However, this speed-up might fail to appear for vectors of reasonable size and sparsity.
For this reason, we study empirically the relative costs of these two algorithms.
In Fig.\ \ref{fig:eval_permGR}, we plot the ratio between the number of Toffoli gates required by Alg.\ \ref{alg:gr-perm} and Alg.\ \ref{alg:gr} (subjected to the optimization strategy in Alg.\ \ref{alg:optimize_angles}) to prepare the same random vectors used to generate the plots in Fig.\ \ref{fig:permGR}.
We find that Alg.\ \ref{alg:gr-perm} performs better than the optimized version of Alg.\ \ref{alg:gr} already at moderate values of $n$ and starting at densities, $d/N$, between $10^{-3}$ and $10^{-2}$.
Unsurprisingly, for larger values of $n$ the transition happens sooner, i.e.\ at larger densities.

\section{Conclusions}
In this paper, we have studied the performance of the Grover-Rudolph algorithm for preparing sparse states. 
We have found that the usual version of the algorithm, see Alg.\ \ref{alg:gr}, has a gate complexity of order $O(dn^2)$.
Here $n$ is the number of qubits needed to encode the vector we want to prepare, $\psi$, and $d$ is the number of nonzero entries in $\psi$ .
Moreover, we have introduced a simple modification of the algorithm which has a gate complexity of order $O(dn)$, Alg.\ \ref{alg:gr-perm}.
This is competitive with the best known algorithms for preparing sparse vectors \cite{gleinig2021efficient, malvetti2021quantum, de2022double, mozafari2022efficient}.
The classical complexity of both Alg.\ \ref{alg:gr} and Alg.\ \ref{alg:gr-perm} is $O(dn)$.
This is better than those of \cite{gleinig2021efficient} and \cite{malvetti2021quantum}, which are $O(nd^2\log d)$ and $O(nd^2)$ respectively, and it's equal to that of \cite{de2022double} and \cite{mozafari2022efficient}.
Ultimately, the decision of which algorithm to use depends on the specific properties of the vectors to prepare. 
The main point of this work is that, when considering sparse vectors, Grover-Rudolph should also be considered as an option. 

Finally, we point out that in both Alg.\ \ref{alg:gr} and Alg.\ \ref{alg:gr-perm} there is space for improvements.
In particular, it would be interesting to consider optimization procedures to reduce the number of controlled rotations and Toffoli gates in Alg.\ \ref{alg:gr}.
We consider one such optimization procedure in App.\ \ref{app:optimization} which shows promising results, for real vectors.
In Alg.\ \ref{alg:gr-perm}, it would be interesting to try to improve the permutation step, both at the level of the classical preprocessing, i.e.\ finding a different suitable permutation, and at the level of the quantum circuit needed to implement the permutation.

\section*{Acknowledgement}
All the results were obtained using Python. The code is available on \href{https://github.com/qubrabench/grover-rudolph}{github.com/qubrabench/grover-rudolph}.
This work was supported by the Quantum Valley Lower Saxony, and the BMBF project QuBRA. Helpful correspondence and discussions with Joshua Ammermann, Lennart Binkowski, Tim Bittner, Domenik Eichhorn, Davide Incalza, Andrei Loțan, Tobias J. Osborne, Anurudh Peduri, Sören Wilkening, and Henrik Wilming are gratefully acknowledged.
\appendix

\section{Optimizing the gates}\label{app:optimization}
\begin{figure}[h!]
    \centering
\includegraphics[scale=0.8]{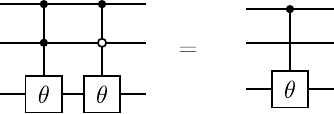}
    \caption{Example of optimization procedure.}
    \label{fig:merging_example}
\end{figure}

We consider a simple optimization strategy in which we merge consecutive gates that have the same rotation angles and phases, and have controls differing only by one bit-flip.
For example, consider the situation depicted in Fig.\ \ref{fig:merging_example}.
The first gate is conditioned on `11' and performs a rotation with an angle $\theta$, while the second gate is conditioned on `10' and executes a rotation with the same angle $\theta$. Given the gates differ in only one control and share the same rotation angles, they can be combined into a single gate, controlled on the first qubit being in `1'.
Notice that this reduction not only affects the total gate count but also leads to gates with one fewer controlling qubit. 

To explain how to perform the merge, we assume again that the angles and phases needed to implement the unitaries $U_k$ are stored in dictionaries $L_k$ as $\{(i_1,\dots, i_k)\text{: }(\theta^{(k)}_{i_1\dots i_k}, \phi^{(k)}_{i_1\dots i_k})\}$.
For every control $(i_1,\dots, i_k)$ in dictionary $L_k$, we loop over its neighbors and check whether the corresponding angles, if present, are the same. 
Here the neighbors are found by flipping one bit, i.e.\ they are neighbors in Hamming distance.
When the angle is the same, we merge the two controls into one.
To do this, we replace the one bit on which the original controls disagreed with `e'.
This simply helps to keep track of which qubits control the rotation.
For example, `010' and `000' would be merged in `0e0'.
We repeat this procedure until no new merging is possible.
Notice that to find neighbors, we don't consider the entries set to `e'.
These steps are summarized in Alg.\ \ref{alg:optimize_angles}.

\begin{algorithm}[h!]
\caption{Optimize Angles}
\label{alg:optimize_angles}
\begin{algorithmic}[1]
\Function{OptimizeAngles}{dictionary D}
\State Merging\_success $\gets$ True \Comment{Flag to mark merging success}
\While{(Merging\_success = True) \& (len(D) $> 1$)}
    \State Merging\_success = \Call{Mergeable}{D}
\EndWhile
\State \Return D
\EndFunction
\State
\Function{Mergeable}{dictionary D}
\For{$k, \theta$ \textbf{in} D}
    \For{$i=0,\dots, \text{len}(k)$}
        \If{$k[i]= $ `e'}
            continue
        \EndIf
        \State $k' \gets$ copy of $k$ with $i$-th entry flipped
        \State $\theta' \gets $ D$[k']$
        \If{$\theta = \theta'$}
        \State Remove $k, k'$ from D
        \State $k'' \gets$ copy of $k$ with $i$-th entry set to `e'
        \State Add $\{k'': \theta\}$ to D
        \State \Return True
        \EndIf
    \EndFor
\EndFor
\State \Return False 
\EndFunction
\end{algorithmic}
\end{algorithm}

The complexity of Alg.\ \ref{alg:optimize_angles} is $O(dn^2\log d)$.
To see this, consider first the function Mergeable.
This has complexity of order $O(dn^2)$, as can it be seen by considering the structure of the two nested for loops.
The first for loop iterates over all the items in the dictionary, whose number is upper bounded by $d$.
The second for loop iterates over all bits in a key, whose number is upper bounded by $n$.
Finally, the operations inside the inner for loop have complexity $O(n)$.
Next we consider the while loop.
This takes at most $\log d$ repetitions, as can be seen by noticing that at any iterations of the while loop, only angles which have been merged in the previous iteration can be further merged.
Hence the number of mergeable angles decreases by at least a factor 2 at every repetition of the while loop.
Therefore, we have at most $O(\log d)$ iterations.
Putting everything together, we arrive to a complexity of order $O(dn^2\log d)$.  

\begin{figure}[h!]
    \centering
     \begin{minipage}{.49\textwidth}
        \centering
        \includegraphics[width=\textwidth]{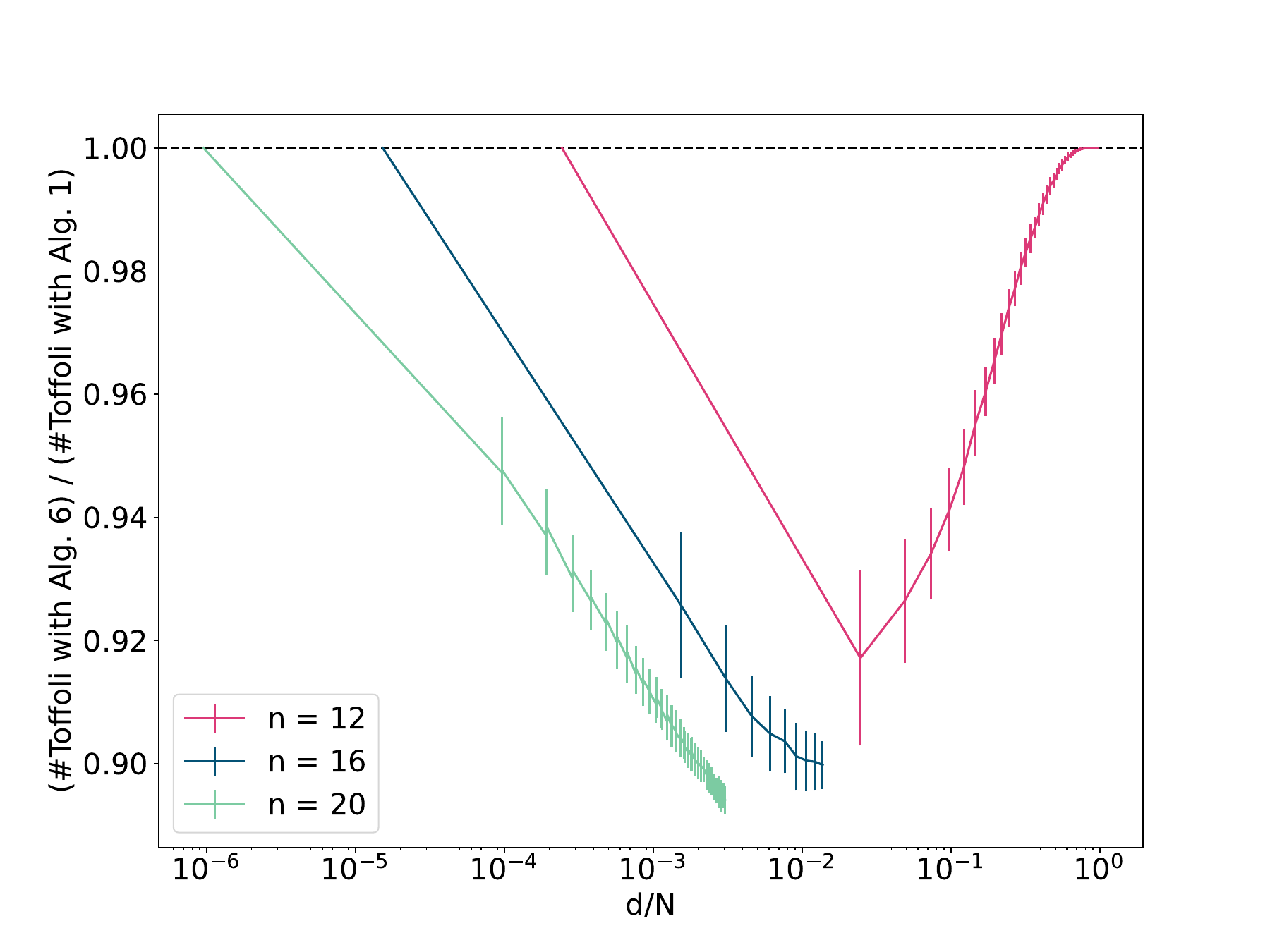} 
    \end{minipage}\hfill
    \begin{minipage}{0.49\textwidth}
        \centering
        \includegraphics[width=\textwidth]{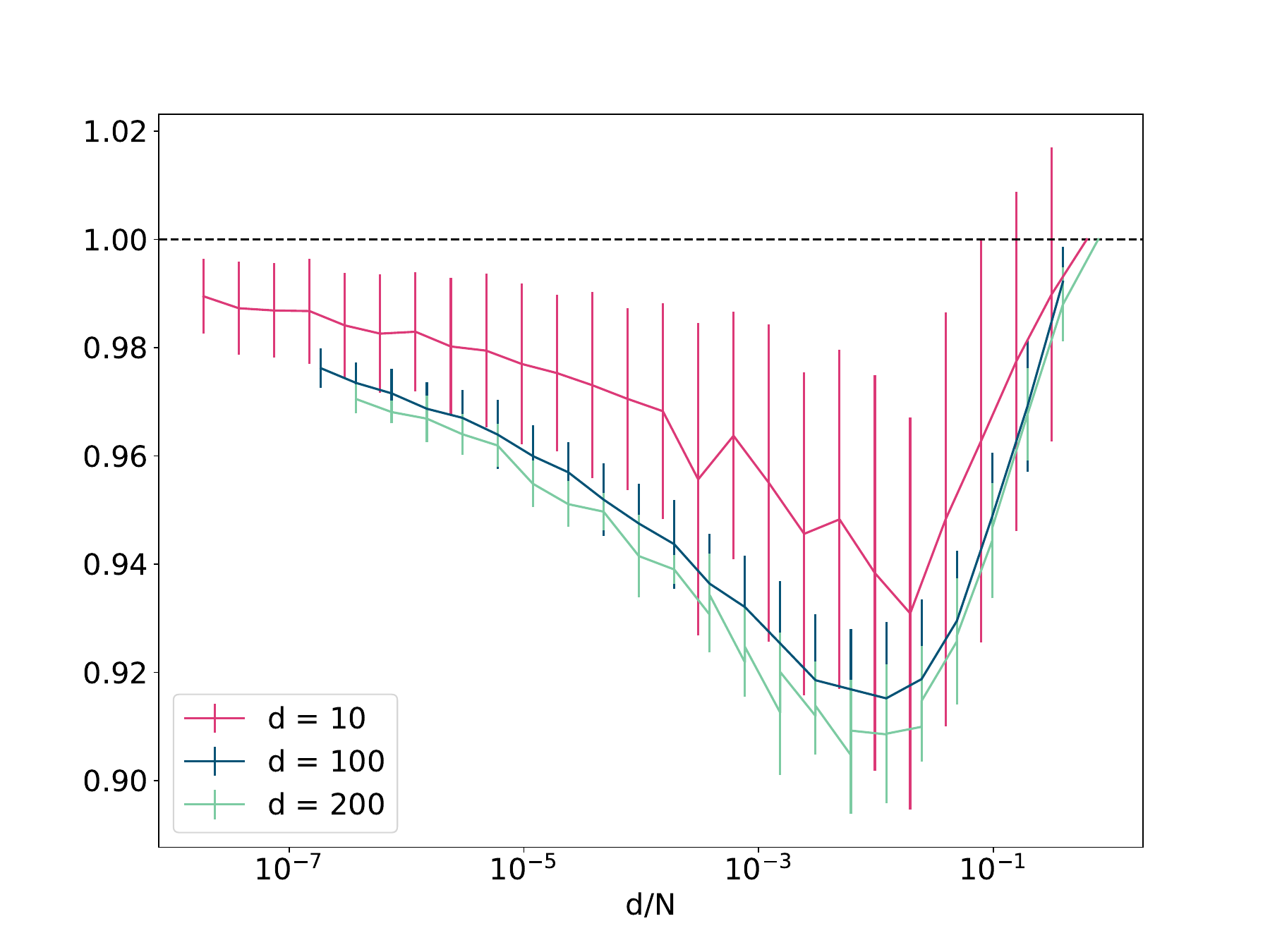}
    \end{minipage}
    \begin{minipage}{0.49\textwidth}
        \centering
        \includegraphics[width=\textwidth]{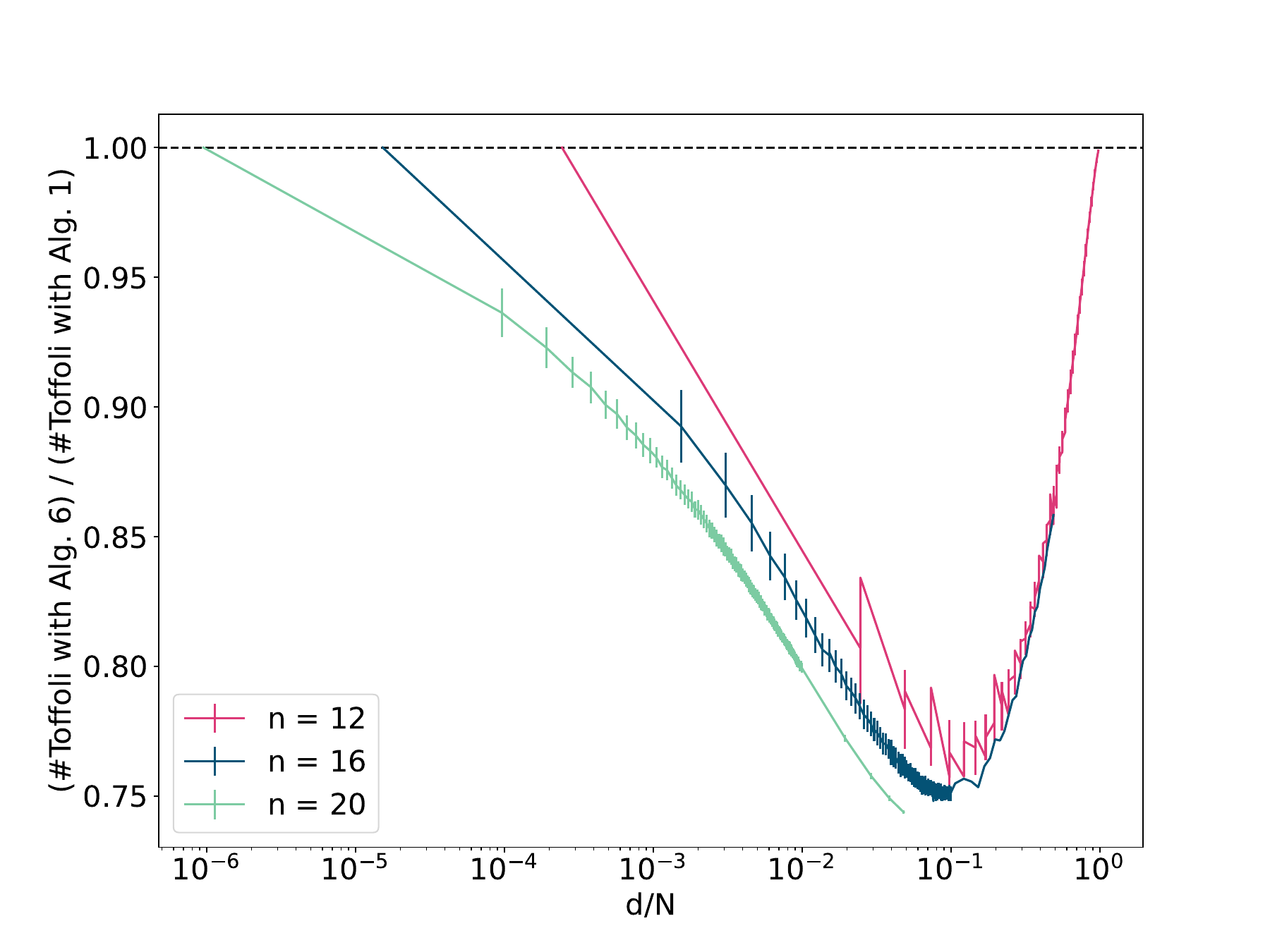}
    \end{minipage}\hfill
    \begin{minipage}{0.49\textwidth}
        \centering
        \includegraphics[width=\textwidth]{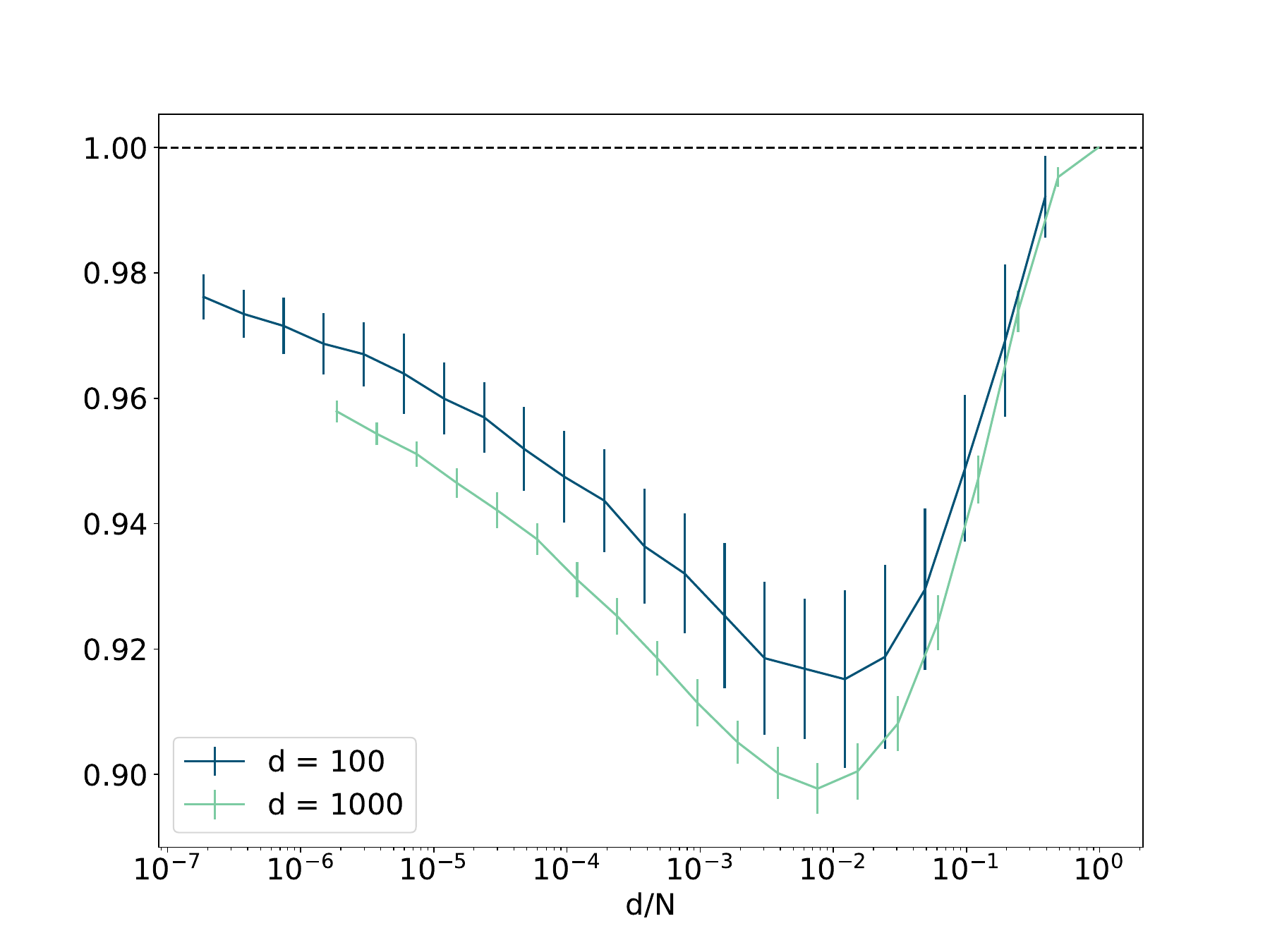}
    \end{minipage}
    \begin{minipage}{0.49\textwidth}
        \centering
        \includegraphics[width=\textwidth]{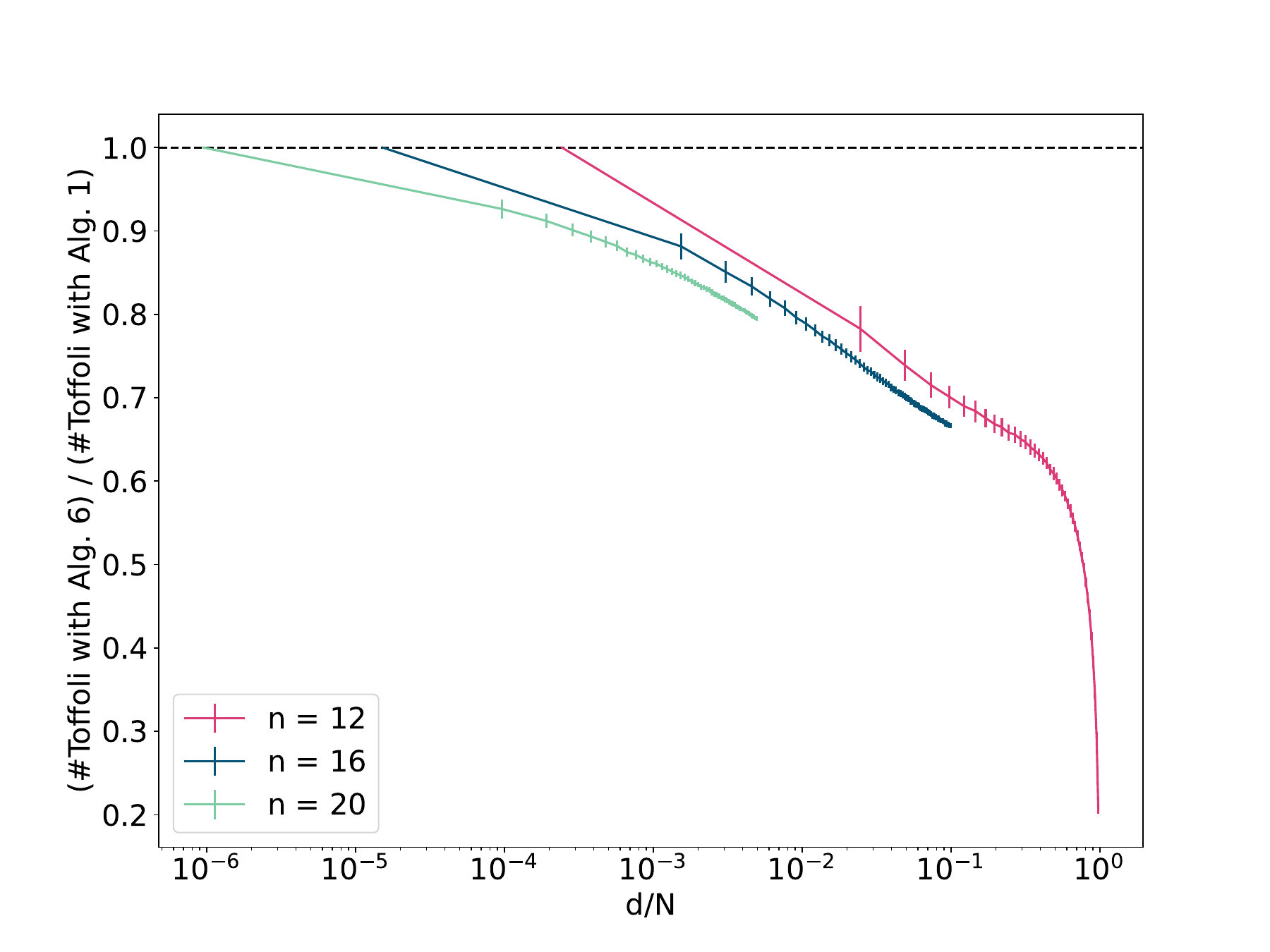}
    \end{minipage}\hfill
    \begin{minipage}{0.49\textwidth}
        \centering
        \includegraphics[width=\textwidth]{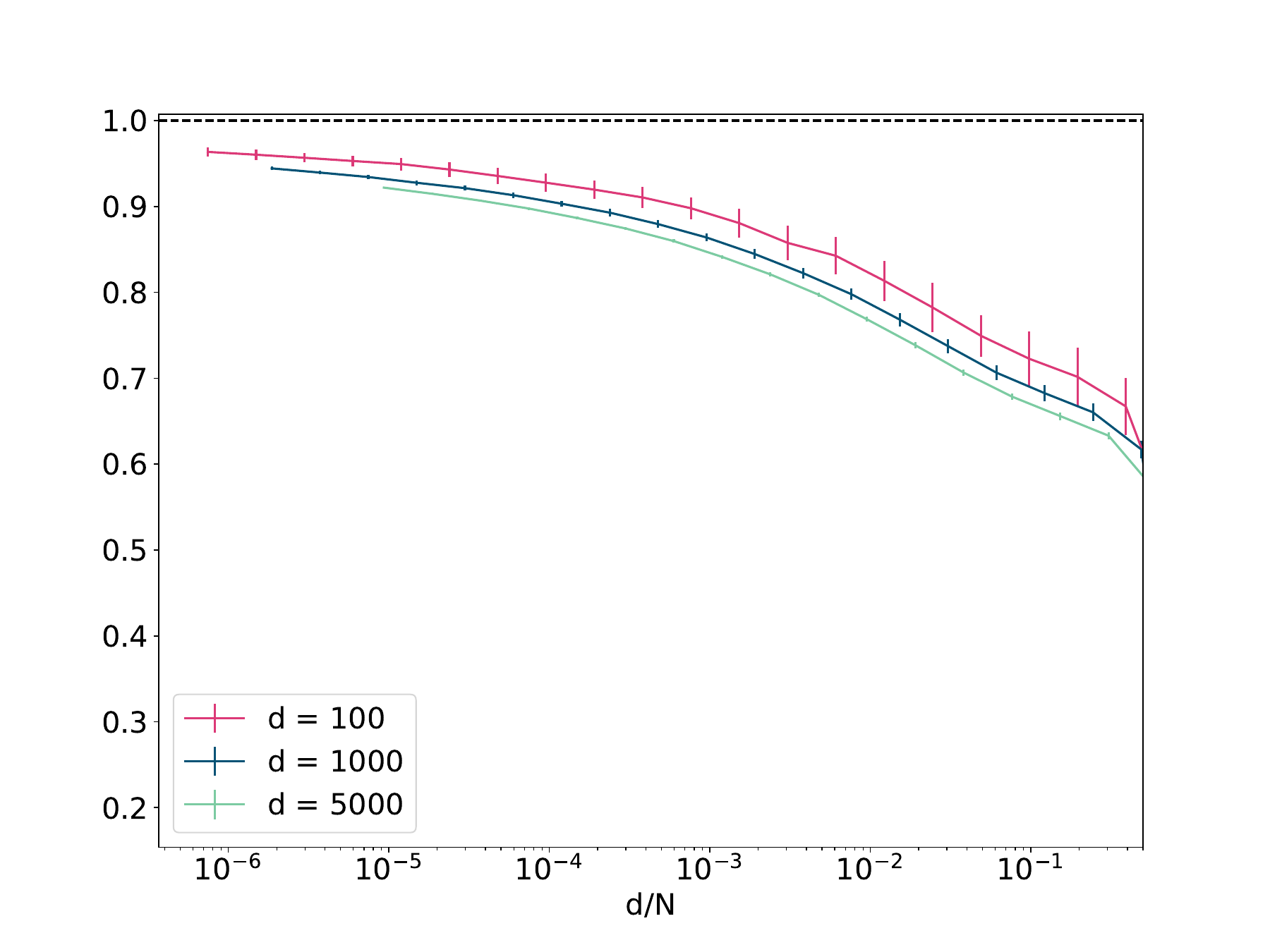}
    \end{minipage}
    \caption{Ratio of number of Toffoli gates needed to run Alg.\ \ref{alg:gr} after and before optimizing the angles using Alg.\ref{alg:optimize_angles} to prepare random states (a), (b), random \textit{real} states (c), (d), and \textit{uniform} states (e), (f) as a function of the density $d/N$ at fixed $n$, and at fixed $d$. We use a log scale on the abscissa.}
    \label{fig:ratio_random_real_uniform}
\end{figure}

It is difficult to understand theoretically how much this optimization reduces the number of required gates.
Therefore, we rely on numerics. We consider the same random vectors used to generate Fig.\ \ref{fig:sparseGR}, we optimize the angles using Alg.\ \ref{alg:optimize_angles}, and compute the ratio between the gates needed to prepare the state before and after optimization.
The results are shown in Fig.\ \ref{fig:ratio_random_real_uniform} (a), (b).
We apply the same strategy further for both real and uniform random vectors and we show the results in Fig.\ \ref{fig:ratio_random_real_uniform} (c) and (d) and Fig.\ \ref{fig:ratio_random_real_uniform} (e) and (f), respectively. 

We find that the optimization of random vectors is only relevant at intermediate values of $d$.
Empirically, the best improvement we observe is around 10\%.
As we have seen, this comes at a classical cost, which is $O(n^2\log{d})$ worse than the one required to find the angles. 
In the case of real vectors, we observe that at densities $d/N \approx 0.1$ the improvement in the gate count ranges from $20\%$ to $25\%$, respectively, making them suitable candidates for this particular type of optimization.
In the case of uniform vectors, the improvement in gate count at moderate values of $d$ ranges from $10\%$ to $35\%$.
For fixed $d$, our optimization approach showcases improvements of up to $40\%$.

In the near-future, quantum costs will be the bottleneck of quantum calculations, hence it is still useful to incur a higher classical complexity cost that will reduce the gate total count, even by a modest amount.

\section{Implementing permutation matrices}\label{sec:permutations}
We present a simple way to implement permutation matrices. 

Given a permutation $\sigma$ of $N$ elements, we want to implement a unitary, which we also denote $\sigma$, that acts as $\sigma\ket{i}=\ket{\sigma(i)}$. For simplicity, we assume that $N=2^n$ for some integer $n$, but this condition can be easily relaxed. First, we classically decompose the permutation in cycles, $\sigma=c_0c_1\dots c_{n_c-1}$. Here $n_c$ is the number of required cycles.

\begin{figure}[H]
    \centering
    \includegraphics[scale=0.8]{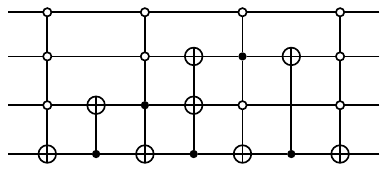}
    \caption{Circuit for $N=8$, $\sigma=(0, 1, 2)$}
    \label{fig:permutation_eg}
\end{figure}

Each cycle is then simple to implement. Let $c=(x_0\;x_1\;\dots\;x_{M-1})$ be a cycle of length $M$, where $x_k$ are $n$-bit strings, and let $\ket{a}$ be an ancilla register that we initially prepare in $\ket{0}$. 
The cycle can be implemented by repeating for $k=0,1,\dots M-1$ the following two operations: 
\begin{enumerate}
    \item We flip the ancilla conditioned on the state of the first $n$ qubits being $\ket{x_k}$. 
    \item Conditioned on the ancilla being in state $\ket{1}$, we map the first $n$ qubits to $\ket{x_{k+1}}$.
\end{enumerate}
To map $\ket{x_k}$ to $\ket{x_{k+1}}$ we can use $\bigotimes_{l=0}^{n-1}X^{\Delta_l}$, where $\Delta = x_k\oplus x_{k+1}$ is the bit-wise difference between $x_k$ and $x_{k+1}$. 
Notice that we are setting $x_M\equiv x_0$. For example, in Figure \ref{fig:permutation_eg}, we show the circuit obtained for cycle $c=(0,1,2)$ and $N=8$.

The steps of the needed to implement a cycle are summarized in Alg.\ \ref{alg:cycle}.
\begin{algorithm}
\caption{Cycle}\label{alg:cycle}
\begin{algorithmic}
\Function{Cycle}{cycle $c=(x_0\;x_1\;\dots\;x_{M-1})$, n-qubit basis state $\ket{y}$}
\State $x_M\gets x_0$
\State Add a qubit ancilla in state $\ket{a}=\ket{0}$
\For{$k =0,1,\dots,M-1$}
\If{$y=x_k$}
\State Flip ancilla, $a\rightarrow a\oplus 1$
\EndIf
\State $\Delta \gets x_k\oplus x_{k+1}$
\If{$a=1$}
\State $\ket{y} \gets (X^{\Delta_0}\otimes X^{\Delta_1}\otimes\dots\otimes X^{\Delta_{n-1}})\cdot\ket{y}$ 
\Comment Map $\ket{x_k}$ to $\ket{x_{k+1}}$
\EndIf
\EndFor
\State \Return $\ket{y}$
\EndFunction
\end{algorithmic}
\end{algorithm}

Finally, we provide a count for the number of gates required to run this algorithm. 
The ancilla flip is a generalized Toffoli gate with $n$ controls, hence the cost of each is 
$2(n-1)\cost_T+4\cost_1+2\cost_{CNOT}$. We also need to include two $X$ gates for every qubit controlled on $0$, resulting in adding $n - \abs{x_k}$ gates.
There will be $M+1$ of these terms. Then each state flip is determined by the number of bits that need to be swapped. More precisely, we need $\abs{x_k\oplus x_{k+1}}$ CNOT gates for each of the $M$ elements of a cycle.

The cost of this algorithm is 
\begin{equation}\label{eq:cost_perm}
\begin{split}
\cost[\text{Cycle}(c)]&=2(M+1)((n-1)\cost_T+2\cost_1+\cost_{CNOT})+2\sum_{k=0}^{M}(n-\abs{x_k})\cost_1\\
&+4\sum_{k=0}^{M-1}\abs{x_k\oplus x_{k+1}}\cost_1+2\sum_{k=0}^{M-1}\abs{x_k\oplus x_{k+1}}\cost_{CNOT}\\
&= O(Mn)\,,
\end{split}
\end{equation}
where $\cost_T$, $C_{CNOT}$, $C_1$ are the costs of Toffoli, CNOT, 1-qubit gate, respectively,  $c=(x_0\;x_1\;\dots\;x_{M-1})$, $\abs{\cdot}$ denotes the Hamming weight of the bit string, and again $x_M\equiv x_0$.
Notice that the algorithm also requires $O(Mn)$ classical operations. 

\bibliographystyle{unsrt}
\bibliography{main}
\end{document}